\documentstyle[preprint,aps,epsf,floats]{revtex}
\begin{document}
\tighten

\def\bfl{{\bbox \ell}}
\def\bull{\vrule height .9ex width .8ex depth -.1ex}
\def\MeV{{\rm MeV}}
\def\GeV{{\rm GeV}}
\def\Tr{{\rm Tr\,}}
\def\nrcpt{NR\raise.4ex\hbox{$\chi$}PT\ }
\def\ket#1{\vert#1\rangle}
\def\bra#1{\langle#1\vert}
\def\ltap{\ \raise.3ex\hbox{$<$\kern-.75em\lower1ex\hbox{$\sim$}}\ }
\def\gtap{\ \raise.3ex\hbox{$>$\kern-.75em\lower1ex\hbox{$\sim$}}\ }
\def\abs#1{\left| #1\right|}
\def\CA{{\cal A}}
\def\CC{{\cal C}}
\def\CD{{\cal D}}
\def\CE{{\cal E}}
\def\CL{{\cal L}}
\def\CO{{\cal O}}
\def\CZ{{\cal Z}}
\def\bvert{\Bigl\vert\Bigr.}
\def\pds{{\it PDS}\ }
\def\ms{MS}
\def\ddq{{{\rm d}^dq \over (2\pi)^d}\,}
\def\ddqm{{{\rm d}^{d-1}{\bf q} \over (2\pi)^{d-1}}\,}
\def\bfq{{\bf q}}
\def\bfk{{\bf k}}
\def\bfp{{\bf p}}
\def\bfpp{{\bf p '}}
\def\bfr{{\bf r}}
\def\dtr{{\rm d}^3\bfr\,}
\def\bfx{{\bf x}}
\def\dtx{{\rm d}^3\bfx\,}
\def\dfx{{\rm d}^4 x\,}
\def\bfy{{\bf y}}
\def\dty{{\rm d}^3\bfy\,}
\def\dfy{{\rm d}^4 y\,}
\def\dfq{{{\rm d}^4 q\over (2\pi)^4}\,}
\def\dfk{{{\rm d}^4 k\over (2\pi)^4}\,}
\def\dfl{{{\rm d}^4 \ell\over (2\pi)^4}\,}
\def\dtq{{{\rm d}^3 {\bf q}\over (2\pi)^3}\,}
\def\dtk{{{\rm d}^3 {\bf k}\over (2\pi)^3}\,}
\def\dtl{{{\rm d}^3 {\bfl}\over (2\pi)^3}\,}
\def\dt{{\rm d}t\,}
\def\frac#1#2{{\textstyle{#1\over#2}}}
\def\darr#1{\raise1.5ex\hbox{$\leftrightarrow$}\mkern-16.5mu #1}
\def\){\right)}
\def\({\left( }
\def\]{\right] }
\def\[{\left[ }
\def\si{{}^1\kern-.14em S_0}
\def\siii{{}^3\kern-.14em S_1}
\def\diii{{}^3\kern-.14em D_1}
\def\dtwiii{{}^3\kern-.14em D_2}
\def\dthiii{{}^3\kern-.14em D_3}
\def\pziii{{}^3\kern-.14em P_0}
\def\poiii{{}^3\kern-.14em P_1}
\def\ptiii{{}^3\kern-.14em P_2}
\def\ipi{{}^1\kern-.14em P_1}
\def\idii{{}^1\kern-.14em D_2}
\def\fm{{\rm\ fm}}
\def\MeV{{\rm\ MeV}}
\def\CA{{\cal A}}
\def\Czzm{ {\cal A}_{-1[00]} }
\def\Cttm{{\cal A}_{-1[22]} }
\def\Ctzm{{\cal A}_{-1[20]} }
\def\Cztm{ {\cal A}_{-1[02]} }
\def\Czzz{{\cal A}_{0[00]} }
\def\Cttz{ {\cal A}_{0[22]} }
\def\Ctzz{{\cal A}_{0[20]} }
\def\Cztz{{\cal A}_{0[02]} }

\def\Ames{ A }  

\newcommand{\eqn}[1]{\label{eq:#1}}
\newcommand{\refeq}[1]{(\ref{eq:#1})}
\newcommand{\eq}{eq.~\refeq}
\newcommand{\eqs}[2]{eqs.~(\ref{eq:#1}-\ref{eq:#2})}
\newcommand{\eqsii}[2]{eqs.~(\ref{eq:#1}, \ref{eq:#2})}
\newcommand{\Eq}{Eq.~\refeq}
\newcommand{\Eqs}{Eqs.~\refeq}

\def\Journal#1#2#3#4{{#1} {\bf #2}, #3 (#4)}

\def\NCA{\em Nuovo Cimento}
\def\NIM{\em Nucl. Instrum. Methods}
\def\NIMA{{\em Nucl. Instrum. Methods} A}
\def\NPB{{\em Nucl. Phys.} B}
\def\NPA{{\em Nucl. Phys.} A}
\def\NP{{\em Nucl. Phys.} }
\def\PLB{{\em Phys. Lett.} B}
\def\PRL{\em Phys. Rev. Lett.}
\def\PRD{{\em Phys. Rev.} D}
\def\PRC{{\em Phys. Rev.} C}
\def\PRA{{\em Phys. Rev.} A}
\def\PR{{\em Phys. Rev.} }
\def\ZPC{{\em Z. Phys.} C}
\def\SJP{{\em Sov. Phys. JETP}}
\def\SJNP{{\em Sov. Phys. Nucl. Phys.}}

\def\FBS{{\em Few Body Systems Suppl.}}
\def\IJMP{{\em Int. J. Mod. Phys.} A}
\def\UJP{{\em Ukr. J. of Phys.}}
\def\CJP{{\em Can. J. Phys.}}
\def\SCI{{\em Science} }


\def\spol{\alpha_{E0}}
\def\qpol{\alpha_{E2}}
\def\Mspol{\beta_{M0}}
\def\Mqpol{\beta_{M2}}


\preprint{\vbox{
\hbox{ NT@UW-99-1}
\hbox{ CALT-68-2201}
}}
\bigskip
\bigskip

\title{$n+p\rightarrow d+\gamma$
in Effective Field Theory}
\author{Martin  J. Savage}  
\address{ Department of Physics, University of Washington,  
Seattle, WA 98915 }
\address{and\ Jefferson Lab., 12000 Jefferson Avenue, Newport News, 
Virginia 23606
\\ {\tt savage@phys.washington.edu} }  
\author{Kevin A. Scaldeferri and Mark B. Wise}
\address{
California Institute of Technology, Pasadena, CA 91125  
\\  {\tt kevin@theory.caltech.edu}
\\  {\tt wise@theory.caltech.edu}}  
\maketitle

\begin{abstract}
The radiative capture process $n+p\rightarrow d+\gamma$
provides clear evidence for meson exchange currents in
nuclear physics.
We compute this process at low energies using a recently developed 
power counting for the effective field theory
that describes nucleon-nucleon interactions.
The leading order contribution to this process
comes from the photon coupling to the nucleon magnetic moments.
At subleading order there are other 
contributions. Among these are graphs where the photon
couples directly to pions, {\it i.e.} meson exchange
currents.
These diagrams are divergent and require the presence of a local
four-nucleon-one-photon counterterm.
The coefficient of this operator is determined by the measured cross
section, $\sigma^{\rm expt} = 334.2\pm 0.5\ {\rm mb}$, for incident
neutrons with speed $|{\bf v}|=2200~{\rm m/s}$.
\end{abstract}

\vskip 2in

\leftline{November 1998}
%
%
%
%
\vfill\eject


\section{Introduction}

The radiative capture $n+p\rightarrow d+\gamma$ is a
classic nuclear physics process where meson exchange currents play a
role.
For protons at rest and incident neutrons, with  speed 
$|{\bf v}| = 2200\ {\rm m/s}$, the cross section for this process
has the experimental value,
$\sigma^{\rm expt} = 334.2\pm 0.5\ {\rm mb}$\cite{CWCa}.
Naively, one expects that an effective range calculation of this
cross section\cite{BLa,Noyes}
would be very close to this.
However, such a calculation gives a value which is approximately $10\%$
smaller than $\sigma^{\rm expt}$.
As first suggested by Brown and Riska\cite{BrRia}, this
discrepancy can at least partly
be accounted for by the inclusion of meson exchange currents.
More recent work by Park, Min and Rho\cite{PMRa} using effective
field theory with Weinberg's power
counting\cite{Weinberg1}
for the nucleon-nucleon interaction, a resonance saturation
hypothesis for the coefficients of some operators 
and a momentum cutoff finds the value
 $\sigma=334\pm 2\ {\rm mb}$. This prediction is
 relatively insensitive to the value of the cut-off and is compatible
with $\sigma^{\rm expt}$.

Recently, a consistent power counting for the nucleon-nucleon interaction has
been established\cite{KSW}. At leading order in Weinberg's scheme,
pion exchange is included in the $NN$ potential and it is interated to
all orders to predict the $NN$ scattering amplitude \cite{Bira}.
 However, the work of
\cite{KSW,LM,DR} shows that iterating the pions without including the
effects of operators with explicit factors of the quark masses or
derivatives does not give a systematic
improvement in the prediction for the $NN$ scattering amplitude. 
Using the power counting of\cite{KSW},
 the bubble chain formed by multiple insertions of the momentum independent
four-nucleon operator gives the leading scattering amplitude for
systems with large scattering lengths.
Higher derivative operators, operators involving insertions of the light quark
mass matrix and pion exchange are of subleading order and are treated in
perturbation theory.
The expansion parameter used in \cite{KSW} is $Q\sim |{\bf p}|, m_\pi$, and
one expands in $Q/\Lambda$ while keeping all orders in $aQ$
 (see also \cite{Kolck}). Here $\Lambda$
is a nonperturbative hadronic scale and $a$ is the scattering length.
At next-to-leading order (NLO) in the $Q$ expansion simple analytic 
expressions can be derived for physical quantities. Various observables
 in the two-nucleon sector have been determined at NLO with this
power counting, such as the electromagnetic form factors and moments of the
deuteron\cite{KSW2}, the polarizabilities of the deuteron\cite{CGSSpol},
Compton scattering cross sections\cite{CGSScompt,Ccompt} and parity violating
observables\cite{SSpv,KSSWpv}.

In this work we compute the cross section for the radiative capture of 
extremely low
momentum neutrons 
$n+p\rightarrow d+\gamma$ at NLO in the effective field
theory $Q$ expansion.
Capture from the $\siii$ channel is suppressed in the expansion
compared to capture from the $\si$ channel.
At zero-recoil, the amplitude for capture from the $\siii$ channel vanishes
since it is simply the overlap of two orthogonal
eigenstates of the strong Hamiltonian.
Hence, at leading order, only the isovector $^1S_0$ capture occurs and
it arises from the isovector
magnetic moment interactions of the nucleons.
Since the amplitudes for capture from the $\si$ and $\siii$ channels do not
interfere, at NLO the cross section comes only from the
amplitude for capture from the $\si$ channel.
At this order there are contributions from a single insertion of four-nucleon
operators with two derivatives,
from a  single insertion of four-nucleon operators with an insertion of the
quark mass matrix, and from the exchange of a
potential pion.  The potential
pion exchange occurs in graphs with a
nucleon magnetic moment interaction and also in graphs where the potential
pion is minimally
coupled to the electromagnetic field. The later contributions
are historically called meson exchange currents.
In addition to these contributions there is also a contribution from a
four-nucleon-one-photon contact interaction.   The coefficient of this operator
has not been previously determined and a major purpose of this paper is
to fix its value.


\section{Effective Field Theory for Nucleon-Nucleon Interactions}

The terms in the effective Lagrange density describing the interactions
between nucleons, pions, and photons can be classified by the number of
nucleon fields that appear. It is convenient to write 
\begin{equation}
{\cal L} = {\cal L}_0 + {\cal L}_1 + {\cal L}_2 + \ldots, 
\end{equation}
where ${\cal L}_n$ contains $n$-body nucleon operators.

${\cal L}_0$ is constructed from the photon field $A^\mu = (A^0, {\bf A})$
and the pion fields which are incorporated into an $SU(2)$ matrix, 
\begin{equation}
\Sigma = \exp\left({\frac{\displaystyle
2i\Pi}{\displaystyle f}}\right)\ \ \ ,\qquad \Pi = \left( 
\begin{array}{cc}
\pi^0/\sqrt{2} & \pi^+ \\ 
\pi^- & -\pi^0/\sqrt{2} 
\end{array}
\right) \ \ \ \ , 
\end{equation}
where $f=131~MeV$ is the pion decay constant. $\Sigma$ transforms under
the global $SU(2)_L \times SU(2)_R$ chiral and $U(1)_{em}$ gauge symmetries
as 
\begin{equation}
\Sigma \rightarrow L\Sigma R^\dagger, \qquad \Sigma \rightarrow e^{i\alpha
Q_{em}} \Sigma e^{-i\alpha Q_{em}} \ \ \ , 
\end{equation}
where $L\in SU(2)_L$, $R\in SU(2)_R$ and $Q_{em}$ is the charge matrix, 
\begin{equation}
Q_{em} = \left( 
\begin{array}{cc}
1 & 0 \\ 
0 & 0 
\end{array}
\right) \ \ \ . 
\end{equation}
The part of the Lagrange density without nucleon fields is 
\begin{eqnarray} {\cal L}_0 &&= {1\over
2} ({\bf E}^2 - {\bf B}^2)  \ +\ {f^2\over 8} \Tr D_\mu \Sigma D^\mu
\Sigma^\dagger \ +\ {f^2\over 4} \lambda \Tr m_q (\Sigma + \Sigma^\dagger)
\ +\ \ldots \ \ \ \ .  \end{eqnarray} The ellipsis denote operators with
more covariant derivatives $D_\mu$, insertions of the quark mass matrix $m_q
= {\rm diag} (m_u, m_d)$, or factors of the electric and magnetic fields.
The parameter $\lambda$ has dimensions of mass and $m_\pi^2 = \lambda (m_u +
m_d)=(137~ {\rm MeV})^2$. Acting on $\Sigma$, the covariant derivative is 
\begin{equation}
D_\mu \Sigma = \partial_\mu \Sigma + ie [Q_{em},\Sigma] A_\mu \ \ \ . 
\end{equation}

When describing pion-nucleon interactions, it is convenient to introduce the
field $\xi = \exp\left(i \Pi/f\right) = \sqrt{\Sigma}$. Under $SU(2)_L
\times SU(2)_R$ it transforms as 
\begin{equation}
\xi \rightarrow L\xi U^\dagger = U\xi R^\dagger, 
\end{equation}
where $U$ is a complicated nonlinear function of $L,R$, and the pion fields.
Since $U$ depends on the pion fields it has spacetime dependence. The
nucleon fields are introduced in a doublet of spin $1/2$ fields 
\begin{equation}
N = \left({ {p \atop n} }\right) 
\end{equation}
that transforms under the chiral $SU(2)_L \times SU(2)_R$ symmetry as $N
\rightarrow UN$ and under the $U(1)_{em}$ gauge transformation as $N
\rightarrow e^{i\alpha Q_{em}} N$. Acting on nucleon fields, the covariant
derivative is 
\begin{equation}
D_\mu N = (\partial_\mu + V_\mu + ie Q_{em}A_\mu )N \, \, , 
\end{equation}
where 
\begin{eqnarray}
V_\mu &&= {1\over 2} (\xi D_\mu \xi^\dagger + \xi^\dagger D_\mu \xi)\nonumber
\  =\
{1\over 2} (\xi \partial_\mu \xi^\dagger + \xi^\dagger \partial_\mu\xi
+ ie A_\mu (\xi^\dagger Q_{em} \xi - \xi Q_{em} \xi^\dagger))
\ \ \ .
\end{eqnarray}
The covariant derivative of $N$ transforms in the same way as $N$ under $%
SU(2)_L \times SU(2)_R$ transformations
({\it i.e.} $D_\mu N \rightarrow U D_\mu N$)
and under $U(1)$ gauge transformations
({\it i.e.} $D_\mu N \rightarrow e^{i\alpha Q_{em}} D_\mu N$).

The one-body terms in the Lagrange density are 
\begin{eqnarray}
{\cal L}_1 & = & N^\dagger \left(i D_0 + {{\bf D}^2\over 2M}\right) N 
+ {ig_A\over 2} N^\dagger  {\bbox \sigma} \cdot 
(\xi {\bf D} \xi^\dagger - \xi^{\dagger} {\bf D} \xi)
N\nonumber \\
& + &  {e\over 2M} N^\dagger
\left( \kappa_0 + {\kappa_1\over 2} [\xi^\dagger \tau_3\xi
  + \xi \tau_3 \xi^\dagger]\right) {\bbox \sigma} \cdot {\bf B} N
+ \ldots,
\label{lagone}
\end{eqnarray}
where $M=939~{\rm MeV}$ is the nucleon mass and
$\kappa _0={\frac 12}(\kappa _p+\kappa _n)=0.4399\ $ and
$\kappa _1={\frac 12}(\kappa _p-\kappa _n)= 2.35294\ $
are the isoscalar and isovector nucleon magnetic moments
in nuclear magnetons. The nucleon matrix element of the axial
current is $g_A=1.25$.

The two-body Lagrange density needed for NLO calculations is 
\begin{eqnarray}
\CL_2 &=&
-\left(C_0^{(\siii)}+ D_2^{(\siii)} \lambda\Tr m_q\right)
(N^T P_i N)^\dagger(N^T P_i N)
\nonumber\\
 & + & {C_2^{(\siii)}\over 8}
\left[(N^T P_i N)^\dagger
\left(N^T \left[ P_i \overrightarrow {\bf D}^2 +\overleftarrow {\bf D}^2 P_i
    - 2 \overleftarrow {\bf D} P_i \overrightarrow {\bf D} \right] N\right)
 +  h.c.\right]
\nonumber\\
& & -\left(C_0^{(\si)}+ D_2^{(\si)} \lambda\Tr m_q\right)
(N^T \overline{P}_i N)^\dagger(N^T \overline{P}_i N)
\nonumber\\
 & + & {C_2^{(\si)}\over 8}
\left[(N^T \overline{P}_i N)^\dagger
\left(N^T \left[ \overline{P}_i \overrightarrow {\bf D}^2
    +\overleftarrow {\bf D}^2 \overline{P}_i
    - 2 \overleftarrow {\bf D} \overline{P}_i
    \overrightarrow {\bf D} \right] N\right)
 +  h.c.\right]
\nonumber\\
&& + \left[e L_1 \ (N^T P_i N)^\dagger (N^T \overline{P}_3 N) B_i
\ -\ 
 e L_2\  i\epsilon_{ijk} (N^T P_i N)^\dagger (N^T P_j N)  B_k
+h.c. \right] ,
\label{lagtwo}
\end{eqnarray}
where $P_i$ and $\overline{P}_i$ are spin-isospin projectors for 
the spin-triplet channel and the spin-singlet channel respectively,
\begin{eqnarray}
& & P_i \equiv {1\over \sqrt{8}} \sigma_2\sigma_i\tau_2
\ \ \ , 
\qquad \Tr P_i^\dagger P_j ={1\over 2} \delta_{ij}
\nonumber\\
& & \overline{P}_i \equiv {1\over \sqrt{8}} \sigma_2\tau_2\tau_i
\ \ \ , 
\qquad \Tr \overline{P}_i^\dagger \overline{P}_j ={1\over 2} \delta_{ij}
\ \ \ .
\end{eqnarray}
The $\sigma $ matrices act on the nucleon spin indices, while the $\tau $
matrices act on isospin indices. The local operators responsible for $S-D$
mixing do not contribute at NLO. 
Terms in ${\cal L}_2$ involving the pion field have been neglected in 
eq.~(\ref{lagtwo}).

The values of the coefficients of the four-nucleon operators in ${\cal L}_2$
depend on the regularization and subtraction scheme that is adopted. 
The power counting of \cite{KSW} is manifest in the power divergence subtraction
scheme, PDS, and we shall use it in this paper. (Momentum subtraction schemes
can also have this power counting \cite{MehStew,Geg}). 
In PDS one works in $D$ dimensions and the poles at both $D=4$ and $D=3$ 
in the loop integrations associated with Feynman diagrams are subtracted. 
The $D=4$ poles are from logarithmic ultraviolet divergences and the $D=3$
poles are from linear ultraviolet divergences.
There is some freedom in
how the Lagrangian is continued to $D$-dimensions. We choose to keep the
Pauli spin matrices three dimensional and continue the derivatives to 
$D$-dimensions. This is similar to the scheme proposed by t'Hooft and Veltman 
\cite{TV} for chiral gauge theories and ends up being convenient since the
$n+p \rightarrow d+\gamma$ amplitude is proportional to the 
antisymmetric Levi-Civita tensor, $\epsilon_{ijk}$.  
At NLO two basic divergent integrals are encountered. The first is
\begin{eqnarray}
\openup3\jot
I_0&\equiv&\left({\mu \over 2}\right)^{4-D} 
\int {{{\rm d}}^{(D-1)}{\bf  q}\over (2\pi)^{(D-1)}}\, 
\({1\over {\bf q}^2 + a^2}\) 
\nonumber\\
&=&  (\sqrt {a^2}~)^{D-3} \Gamma\({3-D\over 2}\)
{(\mu/2)^{4-D}\over  (4\pi)^{(D-1)/2}}\ .
\label{int1}
\end{eqnarray}
$I_0$ has no pole at $D=4$ but does have a pole at $D=3$. Its value in the PDS
scheme is,
\begin{equation}
I_0^{PDS}=\left({1\over 4\pi}\right) (\mu - \sqrt{a^2}).
\end{equation}
The second is the two loop integral,
\begin{equation}
  I_1 \equiv \left({\mu\over 2}\right)^{2 (4-D)}
  \int\ {d^{D-1}{\bf q}\over (2\pi)^{D-1}}{d^{D-1}{\bf l}\over (2\pi)^{D-1}}  
\ {1\over {\bf q}^2+a^2}\ {1\over {\bf l}^2+b^2}\ {1\over ({\bf q}-{\bf l})^2+c^2}.
\end{equation}
$I_1$ has no pole at $D=3$ but does have the pole, $-1/32\pi^2(D-4)$, at
$D=4$. Therefore $I_1$ has the same value in minimal subtraction (MS) as
in PDS \cite{Int},
\begin{equation} 
I_1^{PDS}=I_1^{MS}=-{1\over 16\pi^2}
\left(\log\left({\sqrt{a^2}+\sqrt{b^2}+\sqrt{c^2}\over\mu}\right)
  \ +\ \delta\right),
\label{int2}
\end{equation}
where
\begin{equation}
\delta={1\over 2}\left(\gamma_E-1-{\rm log}\left({\pi \over 4}\right)\right)
\label{scheme}
\end{equation}
and $\gamma_E$ is Euler's constant. Note that because of its logarithmic
divergence, $ [3I_1/(D-1)]^{PDS}=I_1^{PDS}+1/96\pi^2$. 
There is considerable freedom in
the precise way the subtractions are handled. For example, if the
poles in $D=4$ are subtracted with ${\overline {MS}}$ then $\delta =-1/2$.
Finally, we stress that one cannot blindly evaluate the integrals in 
$D$-dimensions and subtract the poles to get the required PDS value. For
example, if $a$ and $b$ are set to zero then $I_1$ has a double pole at
$D=3$. However, this is associated with a logarithmic
 infrared divergence in three 
dimensions, not an ultraviolet divergence, and so it is not subtracted.

Most of the coefficients in ${\cal L}_2$ have been determined.
At NLO the deuteron magnetic moment \cite{KSW2} is 
\begin{eqnarray}
  \mu_d & = & {e\over 2 M} \left(\kappa_p\ +\ \kappa_n\ +\ L_2\ { 2M \gamma
    (\mu-\gamma)^2\over\pi}\right) 
  \ \ \ ,
\end{eqnarray}
where $\gamma=\sqrt{MB}$ with $B=2.225\ {\rm MeV}$ the binding energy of the 
deuteron and $\mu$ is the subtraction point. The coefficient $L_2$ depends on
the subtraction point in such a way that the physical quantity $\mu_d$ is $\mu$
independent. 
The experimental value of the magnetic moment of the deuteron is
$\mu_d=0.85741$ nuclear magnetons and comparing this with the 
prediction above implies that the coefficient $L_2$ 
(renormalized at $\mu=m_\pi$) is,
\begin{eqnarray}
  L_2 (m_\pi) & = & -0.149\ {\rm fm^4}
\ \ \ \ .
\end{eqnarray}
Note that,
\begin{equation}
N^TP_i\sigma_j N= i \epsilon_{ijk} N^TP_kN,
\end{equation}
and so the operator with coefficient $L_2$ in eq.(\ref{lagtwo}) 
is the same as in \cite{KSW2}.

The coefficients of the four-nucleon operators in eq. (\ref{lagtwo})
that don't involve the electromagnetic field have been fixed from 
comparison with experimental
data on $NN$ scattering. We will review this in the following section. 
The only unknown coefficient that contributes at NLO is $L_1$ and it 
will be determined in this work.

\section{S-wave $NN$ Scattering}

The $^1S_0$ $NN$ scattering amplitude, at center of mass momentum $p$,
 has the expansion 
${\cal A}^{(^1S_0)}(p)=\sum_{n=-1}^\infty {\cal A}^{(^1S_0)}_n(p)$, where
${\cal A}^{(^1S_0)}_n(p)$ is of order $Q^{n}$ . At leading order only the 
four nucleon operator with no derivatives need be included and
\begin{equation}
{\cal A}^{(^1S_0)}_{-1}(p)
= { -C_0^{(^1S_0)}\over 1 + C_0^{(^1S_0)} M
 (\mu + ip)/4 \pi}
 \ \ \ .
\label{firsto}
\end{equation}
It is convenient to break the next order contribution into several pieces,
${\cal A}^{(^1S_0)}_0={\cal A}_0^{(I)}+
{\cal A}_0^{(II)}+{\cal A}_0^{(II)}+{\cal A}_0^{(IV)}+{\cal A}_0^{(V)}$, and,
using PDS, ref. \cite{KSW} found,
\begin{eqnarray}
 {\cal A}_0^{(I)} &=& 
-C_2^{(\si)} p^2
\left[ {{\cal A}^{(^1S_0)}_{-1}\over C_0^{(\si)}  } \right]^2
\ \ \ ,
\nonumber\\
 {\cal A}_0^{(II)} &=&  \left({g_A^2\over 2f^2}\right) \left(-1 + {m_\pi^2\over
4p^2} \ln \left( 1 + {4p^2\over m_\pi^2}\right)\right)
\ ,
\nonumber\\
 {\cal A}_{0}^{(III)} &=& {g_A^2\over f^2} 
 \left( {m_\pi M{\cal A}^{(^1S_0)}_{-1}\over 4\pi}
\right) \Bigg( - {(\mu + ip)\over m_\pi}
+ {m_\pi\over 2p} \left[\tan^{-1} \left({2p\over m_\pi}\right) + {i\over 2} \ln
\left(1+ {4p^2\over m_\pi^2} \right)\right]\Bigg)
\ ,
\nonumber\\
{\cal A}_0^{(IV)} &=& {g_A^2\over 2f^2} \left({m_\pi M{\cal A}^{(^1S_0)}_{-1}\over
4\pi}\right)^2 \Bigg(-\left({\mu + ip\over m_\pi}\right)^2
+  i\tan^{-1} \left({2p\over m_\pi}\right) - {1\over 2} \ln
\left({m_\pi^2 + 4p^2\over\mu^2}\right) +{1 \over 6}- \delta\Bigg)
\ ,
\nonumber\\
{\cal A}_0^{(V)} &=& - D^{(\si)}_2 m_\pi^2 
\left[ {{\cal A}^{(^1S_0)}_{-1}\over C_0^{(\si)}  }\right]^2
\ .
\label{secondo}
\end{eqnarray}
Actually, the above expression for ${\cal A}_0^{(IV)}$ is 
slightly different from that in \cite{KSW} because there
some terms that appear above were
 absorbed into a redefinition of $D^{(\si)}_2$.

The scattering length gets contributions from each order in the $Q$ expansion.
To use these results for the scattering amplitude over a region that
includes very low $p$ it is necessary that the the leading
order amplitude give almost the correct the scattering length. 
This can be achieved at NLO by reordering the expansion in the 
following way \cite{MehStew}. Write
\begin{equation}
C_0^{(\si)}=\bar C_0^{(\si)}+\Delta C_0^{(\si)},
\end{equation}
treat $\bar C_0^{(\si)}$ nonperturbatively and $\Delta C_0^{(\si)}$
as a perturbation. Then in eqs. (\ref{firsto}) and (\ref{secondo})
the following changes occur,
\begin{equation}
C_0^{(\si)} \rightarrow \bar C_0^{(\si)}~~,~~D^{(\si)}_2 m_\pi^2 \rightarrow
D^{(\si)}_2 m_\pi^2+\Delta C_0^{(\si)}.
\end{equation}
The coefficient $\bar C_0^{(\si)}$ is no longer independent
of the light quark masses and is chosen so that the physical value of the
scattering length is close to the first term in its $Q$ expansion,
\begin{eqnarray}
{1 \over a^{(\si)}}=&&\left(\mu+{4\pi \over M\bar C_0^{(\si)}}\right)-
{4\pi(D_2^{(\si)}m_{\pi}^2+\Delta  C_0^{(\si)}) \over M  (\bar C_0^{(\si)})^2}
\nonumber \\
&&-{g_A^2M m_{\pi}^2 \over 4 \pi f^2}
\left(\left(\mu+{4\pi \over M\bar C_0^{(\si)}}\right)
{m_{\pi}-\mu \over m_{\pi}^2}+{1 \over 2} {\rm ln}\left({m_{\pi} \over \mu}\right)
+{\delta \over 2}-{1 \over 12}+{\mu^2 \over 2m_{\pi}^2}\right). 
\end{eqnarray}

The subtraction point is arbitrary and the coefficients
$ \bar C_0^{(\si)}$, $C_2^{(\si)}$
 and $D_2^{(\si)}m_{\pi}^2+\Delta C_0^{(\si)}$ depend on $\mu$ in
such a way that ${\cal A}_{-1}^{(\si)}$ and ${\cal A}_0^{(\si)}$
 are independent of $\mu$. These  coefficients 
 are determined from the measured $NN$ phase shift. We 
will only need the first two of them and for these a
fit over the region $7~{\rm MeV}<p<100~{\rm MeV}$ finds, at $\mu=m_{\pi}$
\cite{MehStew}, 
\begin{eqnarray}
\bar C_0^{(\si)}(m_\pi) =-3.529\fm^2\ ,\ 
C_2^{(\si)}(m_\pi) = 3.04\fm^4
\ .
\label{eq:numfitc}
\end{eqnarray}
Discussions of the results of different  fitting procedures can
be found in \cite{KSW,KSW2,MehStew,SteFurn,CohHan}

In the $^3S_1$ channel, identical formulae hold
for the scattering amplitude and scattering length once the
replacement $^1S_0$ $\rightarrow$$^3S_1$ is made for the superscripts. 
However, the fit to
the data is done a little differently. For processes involving the
deuteron it is convenient to constrain $\bar C_0^{(\si)}$ so that
 ${\cal A}^{(\siii)}_{-1}$ gives the correct deuteron binding energy,
$B=2.2255~{\rm MeV}$. This implies that
\begin{equation}
\bar C_0^{(\siii)}(\mu) = -{4 \pi \over M}\left({1 \over \mu-\gamma}\right),
\end{equation}
where $\gamma=\sqrt{MB}$. In this channel a constrained
fit to the $NN$ phase shift yields
\begin{eqnarray}
\bar C_0^{(\siii)}(m_\pi) =-5.708\fm^2\ ,\ 
 C_2^{(\siii)}(m_\pi) =10.8\fm^4\ .
\end{eqnarray}


\section{Cross section for  Radiative Capture}

The amplitude for
the radiative capture of extremely low momentum neutrons $n+p\rightarrow d+\gamma$
has contributions from both the $\si$ and $\siii$ $NN$
 channels. It can be written as
\begin{eqnarray}
\label{eq:matrix}
i{\cal A}(np\rightarrow d\gamma) & = &
e\ X\ N^T\tau_2\ \sigma_2 \  \left[ {\bbox \sigma}\cdot {\bf k}\ 
\ {\bbox\epsilon} (d)^* \cdot {\bbox \epsilon} (\gamma)^*
  \ -\ {\bbox \sigma} \cdot  {\bbox \epsilon} (\gamma)^*\ 
  \ {\bf k}\cdot {\bbox \epsilon} (d)^* 
  \right] N 
\\ \nonumber
& + &
i e\ Y\  \epsilon^{ijk}\ \epsilon (d)^{i*}\   
k^j\  {\bbox\epsilon} (\gamma)^{k*}
\ (N^T\tau_2 \tau_3 \sigma_2 N)
\ \ \ \ ,
\end{eqnarray}
where  $e=|e|$ is the magnitude of the  electron charge, 
$N$ is 
the doublet of nucleon spinors, ${\bbox \epsilon}(\gamma)$ is 
the polarization vector for the photon, ${\bbox \epsilon}(d)$ is the polarization
vector for the deuteron and ${\bf k}$ is the outgoing photon momentum.
The term with coefficient $X$ corresponds to capture from the $\siii$ channel
while the term with coefficient $Y$ corresponds to capture from the $\si$
channel.
For convenience, we define dimensionless variables $\tilde X$ and $\tilde Y$,
by
\begin{eqnarray}
  X & = & i {2\over M} \sqrt{\pi\over\gamma^3}\ \tilde X
  \ \ ,\ \ 
  Y =  i {2\over M} \sqrt{\pi\over\gamma^3}\ \tilde Y
  \ \ \ \ .
\end{eqnarray}
Both $\tilde X$ and $\tilde Y$ have the $Q$ expansions,
$\tilde X = \tilde X_0+ \tilde X_1+...$, and 
$\tilde Y=\tilde Y_0+ \tilde Y_1+...$, where a subscript $n$ denotes a 
contribution of order $Q^n$.
The capture cross section for neutrons with speed $|{\bf v}|$ 
arising from eq.~(\ref{eq:matrix}) is
\begin{eqnarray}\label{eq:sig}
  \sigma & = & {8\pi \alpha \gamma^3 \over M^5 |{\bf v}|}
  \left[ 2 |\tilde X|^2\ +\ |\tilde  Y|^2\right]
  \ \ \ ,
\end{eqnarray}
where $\alpha$ is the fine-structure constant.

\begin{figure}[t]
\centerline{{\epsfxsize=3.0in \epsfbox{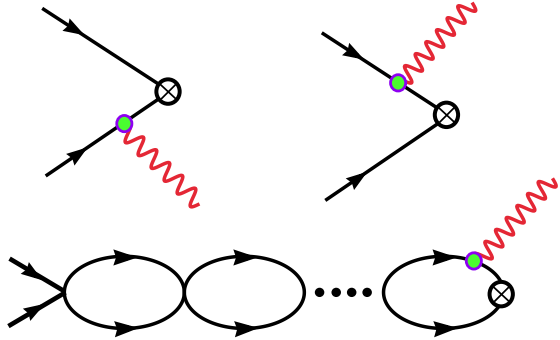}} }
\noindent
\caption{\it Graphs contributing to the amplitude  
  for $n+p\rightarrow d+\gamma$ at leading order in the
  effective field theory expansion. 
  The solid lines denote nucleons
  and the wavy lines denote photons.
The light solid circles correspond to the nucleon magnetic
moment coupling to
the electromagnetic field.
  The crossed circle represents an insertion of the deuteron
  interpolating 
  field which is taken to have $\siii$ quantum numbers.
}
\label{fig:strong}
\vskip .2in
\end{figure}
At leading order,
$\tilde X$ and $\tilde Y$ are calculated from the sum of Feynman diagrams 
in Fig.~(\ref{fig:strong})
and from wavefunction renormalization associated with the deuteron
interpolating
field\cite{KSW2}, giving  
\begin{equation}
  \tilde X_0\ =\
\kappa_0 \ \left( 1\ +\ {\gamma M\over 4\pi}{\cal A}_{-1}^{(\siii)}(0) \right)
\ \ \ \  ,\ \ \ \ 
\tilde Y_0\ =\
\kappa_1 \ \left( 1\ +\ {\gamma M\over 4\pi}{\cal A}_{-1}^{(\si)}(0) \right)
\ \ \ ,
\label{xy}
\end{equation}
where ${\cal A}_{-1}^{(\si)}(p)$ is the leading, order $Q^{-1}$, 
contribution nucleon-nucleon scattering amplitude in the $\si$ channel 
at an center of
mass momentum $p$. The scattering length is related to the nucleon-nucleon
scattering amplitude at zero momentum
\begin{eqnarray}
{\cal A}^{(\si)}(0) = -{4\pi\over M} a^{(\si)}
\ \ \ ,
\label{sl}
\end{eqnarray}
and the experimental value for the $^1S_0$ scattering length is,
$a^{(\si)}=-23.714 \pm 0.013~  {\rm fm}$.
An analogous expression holds in the $\siii$ channel.
At leading order,  
${\cal A}_{-1}^{(\si)}(0) = -4 \pi a^{(\si)}/M$ and
 ${\cal A}_{-1}^{(\siii)}(0) =-4\pi/ M \gamma$.
Using this in eq. (\ref{xy}) gives,

\begin{equation}
  \tilde X_0\ =\ 0
\ \ \ \  ,\ \ \ \ 
\tilde Y_0\ =\
\kappa_1 \ \left( 1\ -\ \gamma a^{(\si)} \right)
\ \ \ ,
\label{leading}
\end{equation}
which implies the radiative capture cross section,
\begin{equation}
\label{eq:leading}
\sigma^{LO}={8\pi\alpha\gamma^5\kappa_1^2 (a^{(\si)})^2 \over |{\bf v}| M^5}
\left(1-{1 \over \gamma a^{(\si)}} \right)^2
\ =\ 297.2\ {\rm mb}
\ \ \  .
\end{equation}
This agrees with the results of Bethe and Longmire~\cite{BLa,Noyes}
when terms in their expression involving the effective range are neglected.
Eq.~(\ref{eq:leading}) is about $10\%$ less than the experimental value,
$\sigma^{\rm expt} = 334.2\pm 0.5\ {\rm mb}$\cite{CWCa}.
Because $ \tilde X_0$ vanishes we only need compute $\tilde Y_1$
to obtain the cross section at NLO.

\begin{figure}[t]
\centerline{{\epsfxsize=3.0in \epsfbox{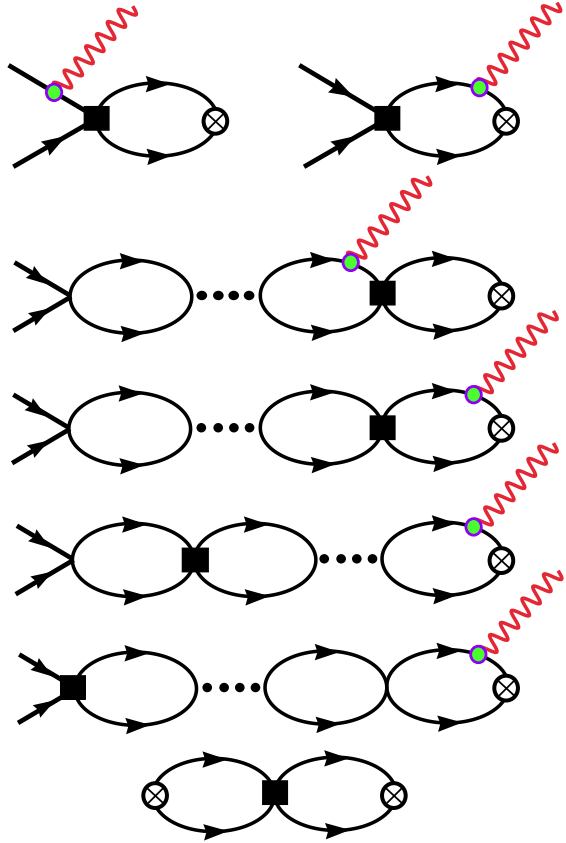}} }
\noindent
\caption{\it Graphs contributing to the amplitude  
  for $n+p\rightarrow d+\gamma$ at subleading order 
  due to the insertion of the $C_2$ and $D_2$ operators.
  The solid lines denote nucleons
  and the wavy lines denote photons.
The light solid circles correspond to the nucleon magnetic
moment coupling of the photon.
The solid square denotes either a $C_2$ operator or a $D_2$
operator. The crossed circle represents an insertion of the deuteron
  interpolating 
  field . The last graph denotes the contribution from wavefunction
renormalization. 
  }
\label{fig:strongsubc2}
\vskip .2in
\end{figure}
%

\begin{figure}[t]
\centerline{{\epsfxsize=3.5in \epsfbox{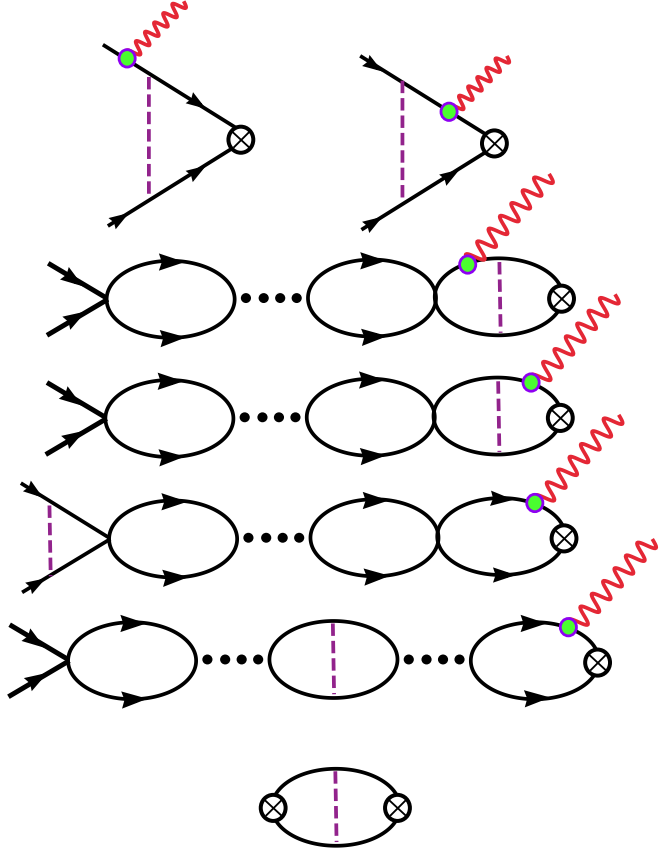}} }
\noindent
\caption{\it Graphs contributing to the amplitude  
  for $n+p\rightarrow d+\gamma$ at subleading order 
  due to the exchange of a potential pion with the photon 
  coupling to the magnetic moment of the nucleons.
  The solid lines denote nucleons
  and the wavy lines denote photons. The dashed line denotes a pion.
The light solid circles correspond to the nucleon magnetic
moment coupling of the photon.
The crossed circle represents an insertion of the deuteron interpolating field.
The last graph denotes the contribution from wavefunction
renormalization. }
\label{fig:strongsubpiB}
\vskip .2in
\end{figure}
%

\begin{figure}[t]
\centerline{{\epsfxsize=3.0in \epsfbox{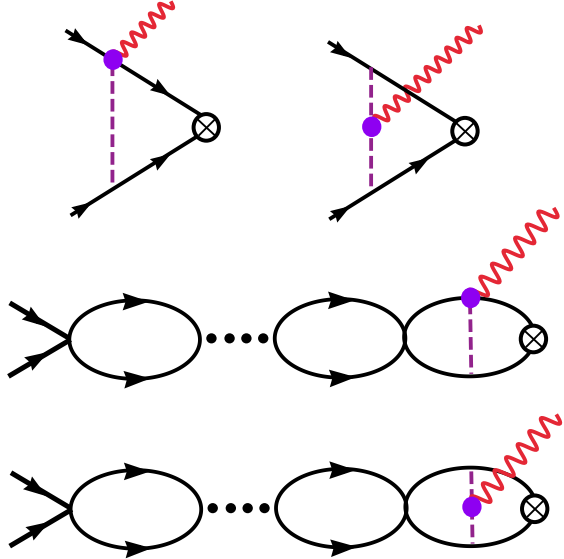}} }
\noindent
\caption{\it Graphs contributing to the amplitude  
  for $n+p\rightarrow d+\gamma$ at subleading order 
  due to meson exchange currents.
  The solid lines denote nucleons
  and the wavy lines denote photons. The dashed line denotes a pion.
The dark solid circles correspond to minimal coupling of the photon.
The crossed circle represents an insertion of the deuteron interpolating 
  field. }
\label{fig:strongsubpiE}
\vskip .2in
\end{figure}
%

\begin{figure}[t]
\centerline{{\epsfxsize=2.0in \epsfbox{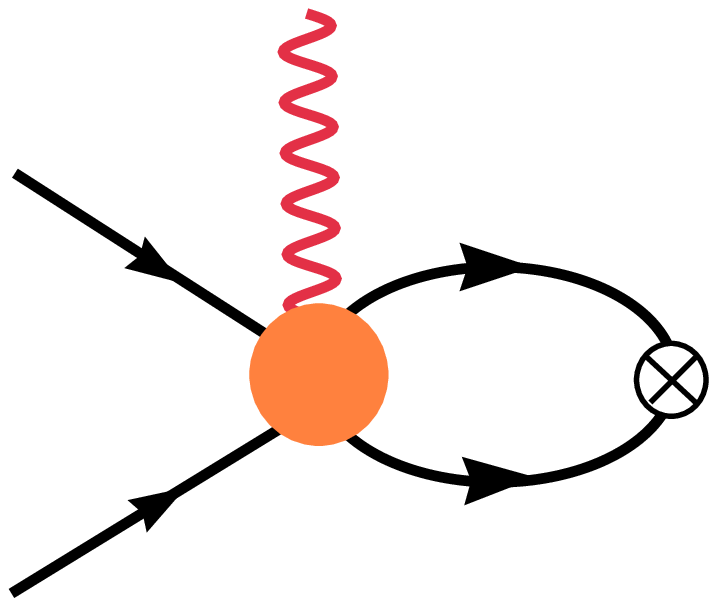}} }
\noindent
\caption{\it Local counterterm contribution to the amplitude  
  for $n+p\rightarrow d+\gamma$ at NLO.
  The solid lines denote nucleons
  and the wavy lines denote photons.
The solid circle corresponds to an insertion of the $L_1$ operator.
  The crossed circle represents an insertion of the deuteron
  interpolating 
  field. }
\label{fig:strongsubpiL1}
\vskip .2in
\end{figure}

At NLO there are contributions from insertions of the $D_2$, $C_2$ operators
and from the exchange of potential pions, with the photon coupling both 
minimally to the pions and to the nucleons via their magnetic moment. 
In addition there is a contribution from the $L_1$ four-nucleon-one-photon
operator.
These can be divided into two categories, those that  build up
the $\si$ scattering amplitude at NLO, ${\cal A}_0^{(\si)}$, and those that do
not. Writing,  $\tilde Y_1=  \tilde Y^{(\rm rescatt)}+\tilde Y^{(\rm C_2)}+
\tilde Y^{(\rm \pi, B)}+\tilde Y^{(\rm \pi, E)}+\tilde Y^{(\rm L_1)}$,
we find the graphs in 
Figs.~(\ref{fig:strongsubc2}), (\ref{fig:strongsubpiB}),
(\ref{fig:strongsubpiE}) and (\ref{fig:strongsubpiL1}) 
give contributions
\begin{eqnarray}
  \tilde Y^{(\rm rescatt)} & = & \kappa_1 {\gamma M\over 4\pi} {\cal
    A}_0^{(\si)}(0)
  \nonumber\\
  \tilde Y^{(\rm C_2)} & = &
  -\kappa_1 {\gamma^2\over 2} \left[ {C_2^{(\si)}+C_2^{(\siii)}\over 
\bar C_0^{(\si)}
      \bar C_0^{(\siii)}}{\cal A}_{-1}^{(\si)}(0)
      \ +\
      {2\over \gamma} {C_2^{(\siii)} (\mu-\gamma)\over \bar C_0^{(\siii)}}
      \left( 1\ +\ {\gamma M\over 4\pi}{\cal A}_{-1}^{(\si)}(0) \right)
      \right]
  \nonumber\\
  \tilde Y^{(\rm \pi, B)} & = & \kappa_1 {g_A^2 M \gamma \over 8\pi f^2}
  \left({m_\pi-2\gamma\over m_\pi+2\gamma}
    \ +\
    {M m_\pi\over 4\pi}{\cal A}_{-1}^{(\si)}(0)
    \left( {m_\pi\over\gamma}{\rm ln}\left(1+2{\gamma\over m_\pi}\right)
      - 2 {m_\pi+\gamma\over m_\pi+2\gamma}\right)\right)
  \nonumber\\
  \tilde Y^{(\rm \pi, E)} & = & {g_A^2 M \gamma ^2 \over 12\pi f^2} 
  \left( {m_\pi-\gamma\over (m_\pi+\gamma)^2}
    \ +\
    { M\over 4\pi}{\cal A}_{-1}^{(\si)}(0)
    \left( { 3m_\pi-\gamma\over 2(m_\pi+\gamma)} +
      {\rm ln}\left({m_\pi+\gamma\over\mu}\right) 
      - {1\over 6} + \delta \right)\right)
  \nonumber\\
  \tilde Y^{(\rm L_1)} & = &  L_1\ \gamma^2\ { {\cal
      A}_{-1}^{(\si)}(0)\over \bar C_0^{(\si)}
     \bar C_0^{(\siii)}}
  \ \ \ .
\label{ys}
\end{eqnarray}
The NLO $\si$ scattering amplitude ${\cal A}_0^{(\si)}$ was given in the
previous section.
Notice that the meson exchange current contribution
$ \tilde Y^{(\rm \pi, E)}$ depends upon the renormalization scale $\mu$.
The graphs contributing to this term have a pole at $D=4$ and
require a subtraction.  This is the reason for the
logarithmic $\mu$-dependence. It is
canceled by the $\mu$-dependence of the constant $L_1$ so that $\tilde Y_1$
is independent of $\mu$.
It is interesting to note that the contribution from the $C_2$ operators,
$\tilde Y^{(\rm C_2)}$, is not $\mu$ independent either.   
This explicit $\mu$-dependence is also canceled by the $\mu$-dependence of $L_1$. 
The renormalization
group equation for the subtraction point dependence of $L_1$ is
\begin{equation}
    \mu {d\over d\mu}
  \left[  { L_1 - {1\over 2}\kappa_1 \left( C_2^{(\si)} +
        C_2^{(\siii)}\right)\over
       \bar C_0^{(\si)} \bar C_0^{(\siii)}}\right]
      ={g_A^2 M^2\over 48\pi^2 f^2}.
\end{equation}
Note that this is quite different from the renormalization group equation,
\begin{equation}
    \mu {d\over d\mu}
    \left[ { L_2 \over (\bar C_0^{(\siii)})^2} \right] 
      =0,
\end{equation}
that $L_2$ satisfies.

There is a NLO contribution that we have not explicitly included, 
a one-loop correction to the magnetic moments of the nucleons that is of
order $m_{\pi}/(4 \pi f^2)$. However, by using the value $\kappa_1=2.35294$,
which follows from the measured values of the neutron and proton magnetic
moments, in $\tilde Y_0$, this effect has been taken into account. Similarly,
using the measured value for $a^{(\si)}$ in  
eq. (\ref{leading}) includes the effects of $\tilde Y^{(\rm rescatt)}$.
 Demanding that the NLO expression for $\tilde Y$ give the measured
cross section implies that,
\begin{equation}
\tilde Y^{(\rm C_2)}+\tilde Y^{(\rm \pi, B)}+\tilde Y^{(\rm \pi, E)}+
\tilde Y^{(\rm L_1)}~=~0.92.
\label{solve}
\end{equation}
In eq. (\ref{ys}) using  $\mu=m_{\pi}$,  $\delta$ 
given by eq. (\ref{scheme}) and 
${\cal A}_{-1}^{(\si)}=-4 \pi a^{(\si)}/M$ yields,
 $\tilde Y^{(\rm C_2)}=0.38$, $\tilde Y^{(\rm \pi, B)}=-0.33$, and 
$\tilde Y^{(\rm \pi, E)}=0.60$. Note that for
 $\tilde Y^{(\rm C_2)}$ there is
a significant cancellation between the two terms in the square brackets
of eq.~(\ref{ys}).
About two thirds of the discrepancy
between the measured cross section and the leading order expression
is made up from the meson exchange current contribution. Most of the rest
comes from $\tilde Y^{(\rm L_1)}$. Eq.~(\ref{solve}) implies that
\begin{eqnarray}
  L_1 (m_\pi) = 1.63\ {\rm fm^4}
  \ \ \  .
\end{eqnarray}
The value of $L_1$ is quite sensitive
to the precise way that the poles at $D=4$ are
handled. For example if $\overline{MS}$ is used ({\it i.e.}
 $\delta=-1/2$) then $\tilde Y^{(\rm \pi, E)}=0.37$ which gives 
$L_1 (m_\pi) = 3.03\ {\rm fm^4}$.

With $L_1$ determined, all the counterterms in the strong and
electromagnetic sector that occur at next-to-leading order in the
effective field theory $Q$ expansion are known.
It is interesting to see that in this framework there is nothing special about
meson exchange currents.  They are simply one of the several contributions at
NLO, occuring along with the strong interaction 
corrections to diagrams where the photon couples to the
nucleon magnetic moments.


\section{Concluding Remarks}

We have computed the cross section for the radiative capture
process $n+p\rightarrow d+\gamma$.
At leading order we recover the effective range theory result (when the terms
involving $r_0$ are neglected)
which is about $10\%$ smaller than the measured
cross section .
At NLO there are contributions from perturbative insertions of the $C_2$
operators, the $D_2$ operators, potential pion exchanges and from a previously
unconstrained four-nucleon-one-photon counterterm with coefficient
$L_1$.
In order to reproduce the  measured cross section,
$\sigma^{\rm expt}$, we find that $L_1 (m_\pi) = 1.65\ {\rm fm^4}$.
In more traditional approaches,
 meson exchange currents are required to explain the value of 
$\sigma^{\rm expt}$. In effective field theory, the meson exchange current
graphs are divergent and require regularization.  As a result, their
contribution to the cross section is not unique and depends upon the choice of
regularization scheme. In addition, a local counterterm is required to absorb 
these divergences and its  value is { \it a priori} unknown and scheme dependent.
Having determined the value of $L_1$ from the radiative capture cross
section, other processes
arising from electromagnetic interactions such as deuteron breakup
$e+ d\rightarrow e^\prime + n + p$ can be computed at NLO. Work on this is
in progress.

\vskip 0.5in

We would like to thank Jiunn-Wei Chen and David Kaplan for several discussions.
This work is supported in part by the U.S. Dept. of Energy under
Grants No. DE-FG03-97ER4014, DE-FG02-96ER40945 and DE-FG03-92-ER40701. 
KS acknowledges support from the NSF under a Graduate Research Fellowship.

\end{document}